\documentclass[pdflatex,sn-vancouver-num,iicol]{sn-jnl}
\usepackage{graphicx}%
\usepackage{multirow}%
\usepackage{amsmath,amssymb,amsfonts}%
\usepackage{amsthm}%
\usepackage{mathrsfs}%
\usepackage[title]{appendix}%
\usepackage{xcolor}%
\usepackage{textcomp}%
\usepackage{manyfoot}%
\usepackage{booktabs}%
\usepackage{algorithm}%
\usepackage{algorithmicx}%
\usepackage{algpseudocode}%
\usepackage{listings}%
\theoremstyle{thmstyleone}%
\theoremstyle{thmstyletwo}%

\theoremstyle{thmstylethree}%

\begin{document}

\title[Mapping Subnational Vulnerability to Inadequate Micronutrient Intake using a Bayesian Small Area Estimation Framework]{Mapping Subnational Vulnerability to Inadequate Micronutrient Intake using a Bayesian Small Area Estimation Framework}

%%=============================================================%%
%% GivenName	-> \fnm{Joergen W.}
%% Particle	-> \spfx{van der} -> surname prefix
%% FamilyName	-> \sur{Ploeg}
%% Suffix	-> \sfx{IV}
%% \author*[1,2]{\fnm{Joergen W.} \spfx{van der} \sur{Ploeg} 
%%  \sfx{IV}}\email{iauthor@gmail.com}
%%=============================================================%%

\author*[1]{\fnm{Sahoko} \sur{Ishida}}\email{sahoko.ishida@cs.ox.ac.uk}
\equalcont{These authors contributed equally to this work.}

\author[2]{\fnm{Mohammed} \sur{Osman}}\email{mohammedaheed.osman@wfp.org}
\equalcont{These authors contributed equally to this work.}

\author[3]{\fnm{Ziyao} \sur{Cui}}\email{richard.cui@duke.edu}

\author[2]{\fnm{Uchenna} \sur{Agu}}\email{uchenna.agu@wfp.org}

\author[2]{\fnm{Emily}
\sur{Becher}}\email{emily.becher@wfp.org}

\author[2]{\fnm{Gabriel} 
\sur{Battcock}}\email{gabriel.battcock@wfp.org}

\author[2]{\fnm{Daniel}
\sur{Hernandez}}\email{daniel.hernandezrive@wfp.org}

\author[4]{\fnm{Duccio} \sur{Piovani}}\email{duccio.piovani@wfp.org}

\author[2,5]{\fnm{Frances}
\sur{Knight}}\email{frances.knight@wfp.org}

\author[1]{\fnm{Seth} \sur{Flaxman}}\email{seth.flaxmna@cs.ox.ac.uk}

\author*[2,5]{\fnm{Kevin} \sur{Tang}}\email{kevin.tang@wfp.org}

\affil[1]{\orgdiv{Department of Computer Science}, \orgname{University of Oxford}, \orgaddress{\street{ Parks Rd}, \city{Oxford}, \postcode{OX1 3QD}, \country{United Kingdom}}}

\affil[2]{\orgdiv{Nutrition and Food Quality Service}, \orgname{World Food Programme}, \orgaddress{\street{Via Cesare Giulio Viola}, \city{Rome}, \postcode{00148}, \country{Italy}}}

\affil[3]{\orgdiv{Department of Computer Science}, \orgname{Duke University}, \orgaddress{\street{Research Drive}, \city{Durham}, \state{North Carolina}, \postcode{27708}, \country{United States}}}

\affil[4]{\orgdiv{Analysis Performance and Planning Division}, \orgname{World Food Programme}, \orgaddress{\street{Via Cesare Giulio Viola}, \city{Rome}, \postcode{00148}, \country{Italy}}}

\affil[5]{\orgdiv{Department of Population Health}, \orgname{London School of Hygiene \& Tropical Medicine}, \orgaddress{\street{Keppel Street}, \city{London}, \postcode{WC1E 7HT}, \country{United Kingdom}}}

%%==================================%%
%% Sample for unstructured abstract %%
%%==================================%%

\abstract{%The Abstract should not exceed 350 words. 
\textbf{Background}: Inadequate dietary micronutrient intake is a significant risk factor for deficiency, which remains a major global health challenge. Nutrition programmes and interventions that aim to address these dietary inadequacies are most effective when targeted to populations at greatest risk. Household Consumption and Expenditure Surveys (HCES) are a widely available source of dietary data, however are often not powered for estimation below the first administrative level, limiting their utility for generating evidence that can inform geographical targeting of interventions. Bayesian Small Area Estimation (SAE) methods offer a framework for generating reliable estimates at finer spatial resolutions.\\
\textbf{Methods}: We constructed household-level indicators of inadequate intake of iron, folate, and vitamin B12 from HCES dietary data. Bayesian SAE models were applied to estimate the prevalence of apparent inadequate intake at the second administrative level. Three approaches were considered: a cluster-level Beta–binomial model and two area-level models (mean smoothing and joint smoothing). Models were evaluated using a HCES survey from Rwanda that supports inference at this spatial scale. All models were implemented in a fully Bayesian manner to propagate uncertainty though all
stages of estimation. Based on simulation results and survey design considerations, appropriate models were selected for application to Senegal and Nigeria.\\
\textbf{Results}: The simulation study in Rwanda showed that the cluster-level Beta–binomial model achieved the strongest performance, while the area-level joint smoothing model provided the most reliable alternative among models that account for survey design. In Senegal, second administrative level estimates for iron, folate, and vitamin B12 captured meaningful subnational variation, substantially reduced uncertainty relative to direct survey estimates, and remained consistent with first administrative level benchmarks after aggregation. Nigeria represents a more challenging setting due to smaller sample sizes at the second administrative level and survey design constraints; nonetheless, modelled estimates reduced extreme uncertainty under direct estimation and showed good agreement with first administrative level direct survey estimates.\\
\textbf{Conclusion}: This study demonstrates that Bayesian SAE frameworks can be applied to routinely collected HCES data to generate more reliable estimates of inadequate micronutrient intake at fine spatial scales, providing decision-makers with the evidence required for localised planning of nutrition programmes and interventions.
}

\keywords{Small area estimation, Population surveys,  Nutrition, Micronutrient intake}  

%%\pacs[JEL Classification]{D8, H51}

%%\pacs[MSC Classification]{35A01, 65L10, 65L12, 65L20, 65L70}

\maketitle

\section{Background}\label{sec: intro}
Micronutrient deficiencies are a major global health concern and are estimated to affect over half of preschool-aged children and over two-thirds of non-pregnant women of reproductive age \citep{stevens2022micronutrient}. These deficiencies are associated with a range of adverse health outcomes \citep{black2013maternal} and can carry significant public health importance. For example, deficiencies in iron, vitamin B12, and folate are common causes of anaemia \citep{lynch2018biomarkers, bailey2015biomarkers,allen2018biomarkers, layden2018neglected} which is estimated to affect 1.8 billion people globally \cite{pasricha2023measuring} and is associated with an increased risk of morbidity and mortality. Furthermore, certain target groups are particularly vulnerable to the deficiency of certain micronutrients due to their increased requirements. For example, pregnant women are particularly vulnerable to folate deficiency which is associated with an increased risk of stillbirth and foetal neural tube defects \citep{bailey2015biomarkers}. Micronutrient deficiencies such as these can arise as a consequence of one of many risk factors including inadequate dietary intake, where usual intake falls below the respective physiological adequacy window that represents a safe range between inadequacy and excess \cite{gibson2005principles, allen2020perspective}. As such, in addition to physiological vulnerability, the prevalence and severity of micronutrient deficiencies are also shaped by broader population-level factors, including poverty, dietary patterns, food systems, and geographic context, which can lead to substantial variation within countries \citep{UNICEFConceptualFramework, osgood2018mapping}. Programme and policy responses can aim to increase micronutrient intake through interventions such as supplementation, food fortification, and dietary diversification.

In many low- and middle-income country (LMIC) settings, governments and partners are increasingly required to make decisions about how to prioritise the allocation of limited resources across competing needs, often with explicit mandates to ensure that the most vulnerable populations are reached first. In this context, subnational estimates of dietary micronutrient inadequacy can play a critical role in supporting programme managers and policymakers to identify populations and geographic areas at highest risk and to guide evidence-based targeting and prioritisation to ensure programmes meet the most vulnerable first\citep{WFPStrategicEvaluation}. Evidence from geospatial analyses of nutrition outcomes has highlighted substantial within-country variation that is not apparent in national or first administrative-level estimates, limiting their usefulness for programme planning. For example, high-resolution mapping of child stunting in Ghana demonstrated how more granular estimates can identify communities with elevated risk and support the targeting of interventions and further investigation\citep{aheto2021geostatistical}.

Household Consumption and Expenditure Surveys (HCES) serve as a practical and routinely collected source of quantitative dietary intake data for estimating household apparent micronutrient intake\citep{zezza2017food, tang2022systematic}. Estimates are derived using household-level consumption data matched with nutrient composition values from Food Composition Tables (FCTs) and are subject to assumptions relating to intra-household food distribution. These estimates should not be conflated with individual-level dietary intake, or biochemical deficiency; however, they serve as a useful proxy for estimating dietary intake of micronutrients and the risk of inadequacy. Despite their utility, HCES are typically designed to be representative only at the first administrative level, limiting their utility for understanding dietary inadequacy at the more local scales where programme decisions are often made. This leaves a critical gap, as HCES can be used to estimate risks of dietary inadequacy but lack the finer-scale resolution needed to target or layer micronutrient interventions effectively. 

Improving the spatial resolution of dietary inadequacy estimates could help identify high-risk geographies, inform the prioritisation of interventions, and guide complementary analyses. In this way, more granular estimates could serve as a decision-support tool to improve the efficiency and equity of nutrition policies and programmes. Small area estimation (SAE) provides a statistical framework to generate reliable subnational estimates in settings where direct survey estimates are unavailable or highly uncertain due to survey design. Direct estimators that rely on survey weights are design-consistent but can exhibit high variance when sample sizes are small, particularly for administrative levels below those for which surveys are designed. SAE methods address this limitation by borrowing strength across areas through statistical modelling. Approaches to SAE can be broadly classified into design-based and model-based frameworks\cite{rao2015small, pfeffermann2002small, pfeffermann2013new, ghosh2020small}. Model-based approaches \cite{Datta2009modelbased} are particularly relevant in settings where data are sparse or unevenly distributed, as they allow information to be combined across areas to stabilise estimates while accounting for uncertainty. 

Model-based SAE methods can be further divided into area-level and unit-level approaches. Area-level models, such as the widely used Fay–Herriot model \cite{fay1979estimates}, operate by smoothing direct estimates across areas with random area specific effects and auxiliary information. However, they typically assume that sampling variances are known, whereas in practice these must often be estimated which leads to additional uncertainty. Extensions to these models include the incorporation of spatially structured random effects \cite{ghosh1998generalized}, approaches that account for unknown sampling variances \citep{kleffe1992estimation, rivest2003mean, wang2003mean}, or that jointly model both the target parameter, such as a mean or prevalence, and its associated sampling variance \cite{you2006small, sugasawa2017bayesian, gao2023spatial}. These developments, particularly joint smoothing, are well suited to applications in LMICs, where survey data are often sparse at subnational levels and limited auxiliary information is available. In contrast, unit-level models operate at the level of individual observations or clusters, such as enumeration areas in household surveys. For discrete outcomes, models such as Beta-binomial \cite{wakefield2020small} are commonly applied to account for overdispersion.

These models can be implemented under different inferential frameworks, including frequentist, empirical Bayes, and fully Bayesian approaches. Fully Bayesian implementations are particularly attractive in data-constrained settings, as they allow uncertainty to be propagated coherently across model components and provide a flexible framework for incorporating hierarchical and spatial structure.

These model-based SAE approaches have been widely applied in disease, mortality and poverty mapping in LMICs \cite{elbers2003micro, mercer2015small, wakefield2019estimating, wakefield2020small, gao2023spatial, gao2024smoothed}  and are increasingly used to inform programmatic decision-making \cite{bedi2007more, worldbank2013povertymapping}. However, their application to estimating the population risk of inadequate micronutrient intake remains limited. This represents an important gap, as dietary assessment poses distinct challenges compared to the anthropometric and mortality outcomes more commonly modelled in this literature.

This study aims to develop a Bayesian small area estimation methodological framework that enables routinely collected household survey data to be used to generate more reliable subnational estimates of dietary micronutrient inadequacy. We focus on fully Bayesian implementations of three approaches: an area-level joint smoothing model, as well as an area-level mean smoothing model, and a cluster-level Beta-binomial model. By improving the spatial resolution of available evidence, estimates can support governments and partners in prioritising and targeting high-risk areas and strengthening the evidence base for nutrition policy and programme planning. The objectives of the study are:
\begin{enumerate}
    \item To evaluate the use of the area-level and cluster-level Bayesian SAE models for estimating inadequate dietary intake of micronutrients, using survey data (Rwanda) that permit reliable design-based estimation at the required administrative level as a reference.
    \item To assess the performance and practical applicability of these models in two settings (Senegal and Nigeria) where surveys are not designed for inference at the required administrative level, and to demonstrate approaches for evaluating modelled estimates in the absence of such reference estimates.
\end{enumerate}
The results of this study are intended to support more data-driven targeting and prioritisation of micronutrient policy and programmes in settings where evidence is currently limited.

\section{Material and methods}
\subsection{Data sources}\label{sec: Data sources}
In this study, we use publicly available HCES data from three countries, Rwanda, Senegal, and Nigeria. Data from Rwanda survey, which was designed for  second administrative (ADM2) level analysis, was used to validate the Bayesian SAE models. Data from two surveys designed for first administrative (ADM1) level analysis (Senegal and Nigeria) with different sampling procedures were then used to demonstrate how the models could be applied to produce ADM2 level estimates in contexts where data were not representative at that level.

\subsubsection{Rwanda EICV7} For Rwanda, we use the Integrated Household Living Conditions Survey (EICV7), 2023-2024, \citep{EICV7}. This is a national household-level survey that is a key source of socio-economic data, including the expenditure and consumption quantities of 148 food items. Data were collected over a 12-month period (October 2023 - October 2024), intending to capture the seasonal variation in household consumption patterns. The survey was designed to be nationally and ADM2-level representative and used a two-stage stratified sampling design, with strata defined by Rwanda’s 30 ADM2 districts. At the first stage, enumeration areas (EAs) were selected from the 2022 Census master sample with probability proportional to size (PPS). A total of 1,674 EAs were drawn, corresponding to 72 per ADM2 in Kigali and 54 per ADM2 elsewhere. Within each ADM2, urban and rural EAs were drawn in proportion to the number of households in each stratum according to the Census frame. At the second stage, nine households per EA were randomly selected for interview, with three additional households listed as replacements to maintain high response rates. A total of 15,054 households were successfully surveyed. This survey design supports statistical inference at the ADM2 level, for both urban and rural strata.

Nutrient composition values for food items listed in the EICV7 were sourced preferentially from the Kenya Food Composition Table 2018 \citep{kenyaFCT}. If unavailable, values were obtained from the West Africa Food Composition Table 2019 \citep{WAFCT}, and finally from the US Department of Agriculture FoodData Central database \citep{USDAfooddata} if the food items were not listed in either of the first two food composition tables.

\subsubsection{Senegal EHCVM 2021/22} For Senegal, we use the Enquête Harmonisée sur le Conditions de Vie des Ménages (EHCVM), 2021-2022, \citep{EHCVM202122}. This survey collected food expenditure and consumption data for 173 food-items. The survey was conducted in two waves, with each comprising roughly half of the total sample. Data for the first wave was collected between November 2021 to January 2022, and data for the second wave was collected between April 2022 and September 2022. This two-wave approach was intended to capture seasonal variation in consumption. The survey is part of the Program for the Harmonisation and Modernisation of Household Living Conditions Surveys in the member states of the West African Economic and Monetary Union (WAEMU), implemented jointly by the WAEMU Commission and the World Bank across 8 countries, including Senegal.  It provides nationally representative household-level data on consumption, expenditure, and food security, designed to produce comparable poverty and living conditions indicators. The household component of this survey, which was used in this analysis, employed a two-stage stratified sampling design, with strata defined by ADM1 and urban–rural classification. EAs were selected with probability PPS from the 2013 General Census of Population and Housing, Agriculture, and Livestock. Within each cluster, twelve households were selected with equal probability, yielding a total target sample of 7,176 households across 598 clusters. The survey design aims to ensure representativeness at the national and ADM1 level.

Nutrient composition values were sourced preferentially from the West Africa Food Composition Table 2019 \citep{WAFCT}, or if unavailable from the US Department of Agriculture FoodData Central Database \cite{USDAfooddata}.

\subsubsection{Nigeria LSS 2018/19} For Nigeria, we use the Living Standards Survey (LSS), 2018-2019, \citep{NLSS}. Like the Rwanda EICV7 and the Senegal EHCVM, this is also a national household survey that collected socio-economic data, including food expenditure and consumption data for 155 food-items. Data were collected over a 12-month period (September 2018 - September 2019) with the intention of capturing seasonal variation in household consumption patterns. The survey was designed to be nationally and ADM1-level representative, with EAs drawn from the National Integrated Survey of Households (NISH2) master sample. In each ADM1, 60 EAs were selected, and within each EA, ten households were chosen with systematic random sampling, which results in an initial sample of 22,200 households. Data collection was spread across 12 months, with household listings updated quarterly to reduce attrition from relocation. The final achieved sample comprised 22,118 households, with a response rate of over 93 percent in every ADM1 and 98 percent overall, although security-related under-coverage occurred in Borno state. The design ensures representativeness at the national, and ADM1-level, with household-level weights calculated based on the selection probabilities of EAs and households, and calibrated to the 2019 total population estimates.

Nutrient composition values for food items listed in the LSS were sourced preferentially from the West Africa Food Composition Table 2019 \citep{WAFCT}, or if unavailable, from the US Department of Agriculture FoodData Central database \citep{USDAfooddata}. 
%%%%%%%%%%%%%%%%%%%%%%%%%%%%%%%%%%%%%%%%%%%%%%%%%%
%%% MIMI INDICATOR %%%%%%%%%%%%%%%%%%%%%%%%%%%%%%%
%%%%%%%%%%%%%%%%%%%%%%%%%%%%%%%%%%%%%%%%%%%%%%%%%%
\subsection{Constructing indicators of inadequate apparent micronutrient intake}

Apparent household micronutrient intake ($I_{h}$) was estimated for folate, vitamin B12 and iron using a nutrient supply model which has previously been described in numerous studies across multiple country contexts \cite{tang2022modeling, tang2025potential, fiedler2012hces}. This model relies on food consumption and household roster data derived from HCES, matched with nutrient composition values derived from national or regional Food Composition Tables. Details of the nutrient supply model used to estimate apparent household micronutrient intake are described fully in Appendix \ref{sec: Indicator construction}

Assumptions were made regarding intra-household food distribution using the Adult Female Equivalent (AFE) approach which is standard practice in dietary analyses using HCES data \cite{weisell2012AME, kalimbira2025AFE}. This approach is fully described in appendix A.1.3 and assumes that food is distributed among household members according to their respective energy requirements. In reality, there may be other factors influencing intra-household food distribution that are not quantified in the survey data, including social, cultural and behavioural dynamics. As a result, apparent micronutrient intake estimates generated from HCES data should be interpreted at the household-level, and cannot be used to infer dietary intake or micronutrient adequacy of individual household members. Consequently, it is not possible to use HCES data to reliably generate insights for specific population groups of interest such as children under 5 or women of reproductive age (WRA). However, household-level apparent intake estimates can still be useful for decision makers as they can be used to construct indicators of inadequate apparent micronutrient intake which can be mapped to characterise geographical differences, thereby supporting subnational targeting of interventions.

The indicators of inadequate apparent micronutrient intake were constructed at the household-level using the apparent intake estimates. For the classification of inadequate intake of folate and vitamin B12, households were categorised as having inadequate intake of the micronutrient if $I_h < H-AR$, where $H-AR$ is the Harmonized Average Requirement of micronutrient intake for a non-pregnant, non-lactating adult female aged 18-24 years old \citep{allen2020perspective}. 

For the classification of the apparent inadequate intake of iron, an alternative approach was taken due to the skewed distribution of individual requirements, primarily due to high variability of individual requirements in children and menstrual iron losses in adolescent and adult females. As such, each household was assigned a probability of inadequacy based on its iron intake (mg/day/AFE), assuming moderate (10\%) bioavailability of dietary iron \citep{IronRequirements}. Households with probability of inadequacy greater than 50\% were classified as inadequate, and others classified as having adequate intake of iron. 

These constructed indicators are based on apparent dietary intake of micronutrients based on reported food consumption, and therefore represent the household-level risk of inadequate micronutrient intake. It is important to not conflate this with true deficiency, which would require measurement of serum biomarkers.

These household-level indicators were then used as the binary outcome in Bayesian SAE models. We denote this variable by $y_{h}$ where $y_{h} = 1$ if the household $h$ is classified as having inadequate micronutrient intake and $y_{h} = 0$ otherwise. 

\subsection{Small area estimation}\label{sec: SAE}
Following the construction of household-level indicators of inadequate micronutrient intake, the next step was to estimate their prevalence at the second administrative level, denoted by $p_\ell$, for each area $\ell = 1,\ldots, L$. 
While the indicator itself was derived using established methods, producing reliable estimates for these areas in settings with limited sample sizes required additional statistical modelling. In this study, we applied SAE methods to generate precise subnational estimates of micronutrient inadequacy. 

As a design-based benchmark, area-level prevalence can be estimated using the direct weighted estimator, known as H\'{a}jek estimator \citep{hajek1971discussion}, given by
\begin{equation}\label{eq: direct estimate}
    \hat{p}^\text{direct}_\ell= \frac{\sum_{h\in\mathcal{S}_\ell}w_h y_h}{\sum_{h\in{\mathcal{S}_\ell}}w_h}
\end{equation}
where $w_h$ is the survey weight for household $h$ and $\mathcal{S}_\ell$ is the set of household indices that belong to area $\ell$. 
We applied model-based SAE methods to obtain stabilised estimates of $p_\ell$, considering both area-level and cluster-level models within a fully Bayesian framework.

\subsubsection{Area-level joint smoothing model}\label{sec: area level model}
We considered an area-level modelling framework based on the Fay--Herriot (FH) model, which treats the direct estimates as noisy observations of the true area-level quantities. Specifically, for each area $\ell$, the sampling model is given by
\begin{equation}\label{eq: gaussian sampling model}
    \hat{p}_\ell^\text{direct} \mid p_\ell, V_\ell \sim N(p_\ell, V_\ell)
\end{equation}
where $p_\ell$ denotes the finite population proportion in area $\ell$, and $V_\ell = \mathrm{Var}(\hat{p}_\ell^\text{direct})$ denotes the sampling variance of the direct estimator.

In practice, $V_\ell$ is unknown and replaced by an estimate $\hat{V}_\ell$, but standard FH models does not incorporate the uncertainty in this estimate. To account for this additional source of uncertainty, we adopted a Bayesian joint smoothing approach proposed by \citet{gao2023spatial} that models both the direct estimates and their sampling variances. 
Their approach combines a spatially structured linking model for the means with a variance smoothing model that accounts for spatial correlation and heterogeneity in the sampling variances.
In what follows, we briefly outline the mean and variance smoothing models, referring the reader to \citet{gao2023spatial} for full technical details.

\paragraph{Mean smoothing}
We follow \citet{gao2023spatial} and model the finite population area-specific proportion $p_\ell$ in  Eq.\eqref{eq: gaussian sampling model} by
\begin{equation}\label{eq: area-level logit smoothing}
    \text{log}\frac{p_\ell}{1-p_\ell} = \beta_0 + u_\ell.
\end{equation}
We note that in the original work by \citet{gao2023spatial}, there is a term modelling the effect of area-level covariates. For this study, we did not use the area-specific covariates; hence, we have the constant term $\beta_0$ and an area-level random effect term $u_\ell$. 

To account for correlation among neighbouring areas, Besag York Molli\'e (BYM) model \citep{besag1991bayesian} was used, which captures both unstructured random effects and structured random effects. More specifically, we used the reparametrisation known as the BYM2 model \citep{riebler2016intuitive}, which has the form:
\begin{equation}\label{eq: BYM2}
    u_\ell = \sigma_u\left(\sqrt{1-\phi} u_{1,\ell} + \sqrt{\phi}\frac{u_{2,\ell}}{\alpha}\right).
\end{equation}
This is the mixture of the unstructured component $u_{1,\ell}$ and structured component $u_{2,\ell}$, where the former is assumed to follow independent standard normal distribution, $u_{1,\ell} \sim N(0, 1)$, and the latter is assigned an intrinsic conditional autoregressive (ICAR) prior \citep{besag1995conditional}, 
\begin{equation}\label{eq: ICAR}
u_{2,\ell}| \boldsymbol{u}_{2,-\ell} \sim N(\frac{\sum_{k\in\partial_\ell}u_\ell}{m_\ell},\frac{1}{m_\ell})
\end{equation}
where  $\partial_\ell$ denotes the set of neighbours for the areas $\ell$,  $m_\ell = |\partial_\ell|$ and $\boldsymbol{u}_{2,-\ell}= \{u_{2,k}: k\neq \ell\}$ so that $u_{2,\ell}$ is normally distributed with mean equal to the average of its neighbours and variance inversely proportional to the number of neighbours. \{This prior encodes the assumption that neighbouring areas are more likely to have similar underlying latent risks than areas that are farther apart, thereby borrowing strength across adjacent areas while permitting local deviations through the unstructured component. However, if neighbouring areas differ substantially in their latent risk and the data provide limited information, the resulting estimates may be oversmoothed toward neighbouring values. The extent to which the overall spatial effect is governed by the structured ICAR component versus the unstructured component is determined by the mixing parameter $\phi$.
This parameter $\phi$ ranges from $0$ to $1$ and measures the proportion of the marginal variance explained by the spatially structured ICAR component. The scaling factor $\alpha$ standardises the structured random effect so that $\sigma_u$ and $\phi$ are interpretable irrespective of the neighbourhood structure; see \citet{riebler2016intuitive} for the technical detail.

\paragraph{Variance smoothing}
The variance estimates $\hat{V}_\ell$ is assumed to follow chi-square distribution 
\begin{equation}\notag
    \frac{d_\ell \hat{V}_\ell}{V_\ell} | d_\ell, V_\ell \sim \chi^2_{d_\ell}
\end{equation}
where the true sampling variance $V_\ell$ is modelled by
\begin{equation}\notag
    \log{(V_\ell)} = \gamma_0 + \gamma_1\log(p_\ell(1-p_\ell))+\gamma_2\log(n_\ell) + \tau_\ell
\end{equation}
and $\tau_\ell\sim_{iid}N(0,\sigma_\tau^2)$. Here $n_\ell$ is the sample size for area $\ell$. The degree of freedom $d_\ell$ is determined by the survey design and how $\hat{V}_\ell$ is computed. Details on how sampling variances and the corresponding degrees of freedom were estimated in this study are provided in Appendix \ref{apx: sampling variance}.
For comparison, we will also fit the model without variance-smoothing, i.e., a mean-smoothing-only model. 

\subsubsection{Cluster-level Beta-binomial model}\label{sec: cluste level model}
 One of the most widely used unit-level models for this type of survey is the Beta-binomial model \citep{wakefield2020small}. In this work, we chose the unit to be cluster, which corresponds to EAs, hence used the cluster-level Beta-binomial model.
Let $\mathcal{S}_c$ denote the index set of households belonging to cluster $c$. Define $y_c = \sum_{h \in \mathcal{S}_c} y_h$ as the number of households in cluster $c$ classified as having inadequate apparent micronutrient intake, and $n_c = |\mathcal{S}_c|$ as the number of sampled households in cluster $c$. We considered the following unit-level model:
\begin{equation}\label{eq: beta-binomial sampling} 
y_c | n_c, r_c \sim \text{BetaBinomial}(r_c, n_c, \lambda)
\end{equation}
where $r_c$ is the shared risk of inadequate micronutrient intake for cluster $c$, and $\lambda$ is an overdispersion parameter. The risk $r_c$ is modelled similar to Eq.\eqref{eq: area-level logit smoothing} with 
\begin{equation}\label{eq: beta-binomial ybut level logit smoothing} 
    \log{\frac{r_c}{1-r_c}} = \beta_0 + \beta z_c + u_{\ell[c]}.
\end{equation}
The area-level random effect is assumed to follow the BYM2 model as described in Eq.\eqref{eq: BYM2} and Eq.\eqref{eq: ICAR}, and $z_c$ is a stratification variable indicating if the cluster is in an urban area. Similarly to the area-level models considered in this paper, we did not use any other cluster-level covariates. The prevalence of insufficient micronutrient intake in area $\ell$ for urban clusters and rural clusters is
\begin{align}
    {p}^{\text{urban}}_\ell &= \frac{1}{1+\exp{\left(-(\beta_0 +\beta + u_\ell)\right)}}, \notag \\
    {p}^{\text{rural}}_\ell &= \frac{1}{1+\exp{\left(-(\beta_0 + u_\ell)\right)}}\notag
\end{align}
respectively. To obtain the aggregated prevalence for area $\ell$, we take the weighted average of the two by 
\begin{equation}\label{eq: betabinomial area level estimate}
    {p}_\ell = q_\ell {p}^{\text{urban}}_\ell + (1-q_\ell) {p}^{\text{rural}}_\ell
\end{equation}
where $q_\ell$ is the proportion of the urban clusters in area $\ell$. 

\subsubsection{Full Bayesian Implementation}
We adopted a fully Bayesian approach by specifying prior distributions for all model parameters. Specifically, we assign independent normal priors with variance $5$ to the regression coefficients $\beta$ and the intercept $\beta_0$. For the parameters $\gamma_0$, $\gamma_1$, and $\gamma_2$, we follow the prior specifications outlined in \cite{gao2023spatial} and assign $N(0,1)$, $N(1,0.5)$ and $N(-1,0.5)$ respectively. 
A penalised complexity (PC) prior \citep{riebler2016intuitive, simpson2017penalising, sorbye2017penalised} is placed on the scale parameter $\sigma_u$ to provide regularisation and to guard against over-fitting. More specifically, we set the prior so that $P(\sigma_u>1) = 0.01$. We used Beta distribution for the prior on the mixing parameter $\phi$. 
Model fitting was carried out in the probabilistic programming language \texttt{Stan} \citep{Stan}, using the No-U-Turn Sampler (NUTS) for posterior inference.

\subsection{Simulation study using the Rwanda EICV7 survey}\label{sec: rwanda experiment}
We assessed the performance of the SAE models using a sub-sampling-based evaluation framework applied to the Rwanda EICV7 survey, which is designed to produce reliable estimates at the ADM2 level. This approach constitutes an internal validation setting, in which sub-samples are drawn from a common parent survey to emulate a typical survey that is designed for ADM1-level inference. We constructed sub-sampled datasets by selecting a limited number of EAs per ADM1, stratified by urban and rural residence, as described in Section~\ref{sec: experiment seup}. Because the sub-samples are derived from the same underlying survey, they inherit common features of the sampling frame, measurement processes, and spatial structure. As a result, the evaluation reflects model performance under controlled conditions and may represent a more favourable setting than fully independent simulation studies. The results are therefore interpreted primarily as informing model choice in relation to survey design, rather than establishing a universal ranking across contexts.

In this validation study, we focused primarily on iron and vitamin~B\textsubscript{12}, given their sufficient prevalence and spatial variability in Rwanda to enable meaningful comparison between direct and model-based estimates. In contrast, folate inadequacy is rare and exhibits limited spatial heterogeneity in this setting.

\begin{figure*}[t]
    \centering    \includegraphics[width=0.99\linewidth]{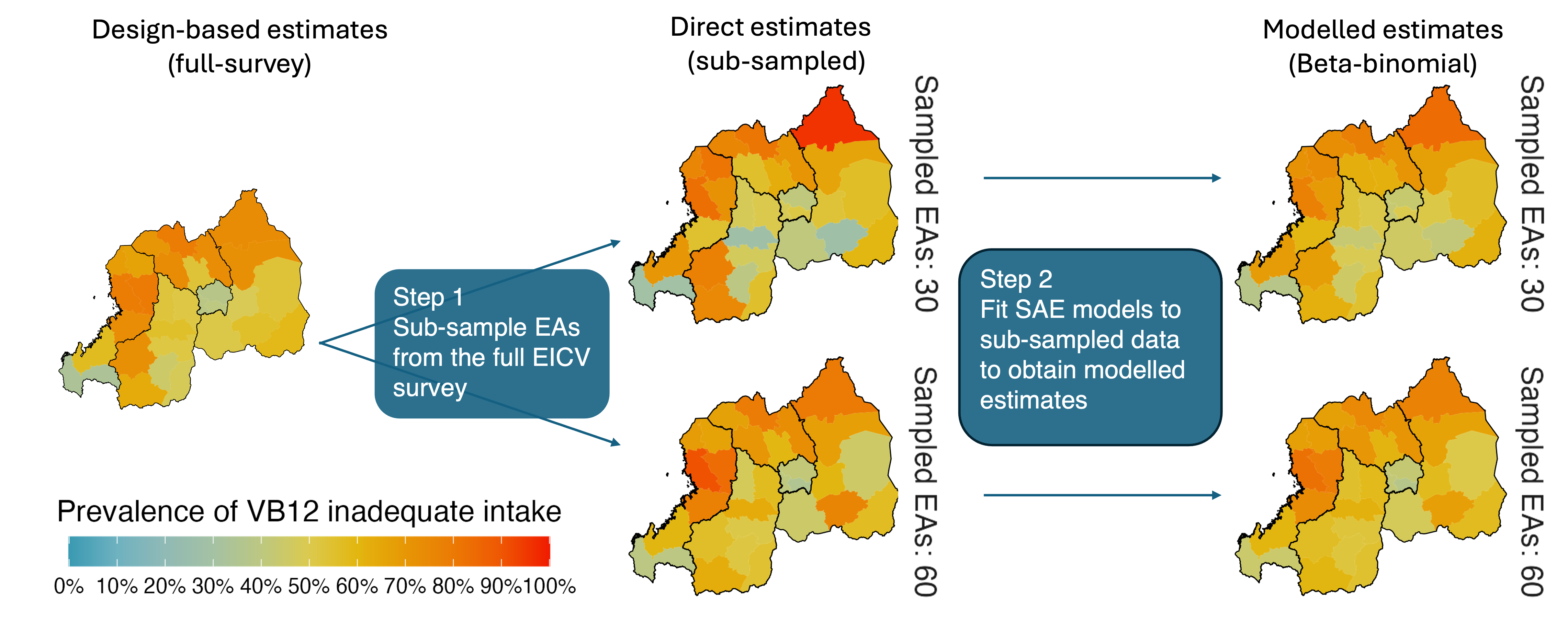}
    \caption{Illustration of the validation study design. Step 1: Enumeration Areas (EAs) are sub-sampled from the full EICV survey to emulate ADM1-representative survey designs with varying sampling intensities (30 or 60 EAs per ADM1). Step 2: SAE models are fitted to each sub-sampled dataset to generate ADM2-level modelled estimates, which are compared with gold-standard direct estimates from the full EICV survey. The maps shown illustrate one simulation example for inadequate vitamin $\text{B}_{12}$ intake with Beta-binomial model; the experiment was repeated 500 times.}
    \label{fig:validation illustration}
\end{figure*}
\subsubsection{Design}\label{sec: experiment seup}
In each simulation replicate, a subset of EAs was drawn from the full dataset to mimic an ADM1-representative survey, where each EA is randomly sampled with PPS. Two sampling intensities were considered to reflect different levels of data sparsity: one with 60 EAs per ADM1 and another with 30 EAs per ADM1. The number of EAs sampled per ADM2 vary across replicates; in the lower-intensity scenario, some ADM2 regions may contain very few or no sampled EAs, which creates a more challenging estimation setting. Sampling was conducted with proportional allocation across urban and rural strata within each ADM1, following the original EICV design in which EA allocation within each ADM2 is proportional to the urban–rural composition. Each replicate sample was treated as an independent pseudo-survey, and the procedure was repeated $500$ times. See Appendix~\ref{apx: subsampling procedure} for the technical details of the sub-sampling procedure.

For each of the 500 sub-sampled datasets, we computed design-based direct estimates at the ADM2 level using survey weights to account for survey design. Model-based estimates were obtained by fitting Bayesian SAE models to the same sub-sampled datasets. We considered two area-level models described in Section~\ref{sec: area level model}: a mean smoothing model, where spatial random effects were specified using a BYM2 prior, and a joint smoothing model, which additionally applied smoothing to the estimates of the sampling variance. We also applied the cluster-level Beta-binomial model described in Section~\ref{sec: cluste level model} and computed the ADM2-level estimates using Eq.\eqref{eq: betabinomial area level estimate}. These two steps are illustrated in Figure~\ref{fig:validation illustration}. The urban-rural proportion at ADM2-level required for the Beta-binomial model is available in the methodology note of the EICV survey \citep{NISR2025EICV7}. 

{Rwanda differs from Senegal and Nigeria in many respects, including population size, geographic extent, and administrative structure. Given these differences, the validation study was not intended to reproduce each application setting exactly; rather, it was designed to evaluate model behaviour under levels of ADM2 sampling sparsity relevant to the household surveys in our application. The two simulation settings were therefore selected to span the range of ADM2 sampling sparsity encountered in these surveys. Rwanda and Senegal also share similar two-stage cluster sampling designs, with EAs selected using PPS sampling. The higher-intensity setting (60 EAs per ADM1) resulted in ADM2-level EA counts that are broadly comparable to those observed in Senegal (3–35 sampled EAs per ADM2). Likewise, the lower-intensity setting (30 EAs per ADM1) produced substantially sparser ADM2 coverage, corresponding more closely to the level of sampling sparsity observed in Nigeria, where many ADM2 units contain between 0 and 10 sampled EAs. Nigeria nevertheless remains a substantially more challenging application because of its much larger number of ADM2 units and the very limited survey information available within many of these areas.
}

\subsubsection{Evaluation}
The performance of the models was evaluated by comparing their ADM2-level estimates with the design-based direct estimates derived from the full ADM2-representative EICV dataset. Although these full-survey estimates still include sampling variability, their precision was high. Precision was assessed using the coefficient of variation (CV), defined as the ratio of the standard error to the estimate. In the full EICV survey, most CVs were below 16.6\% and all were well below 33.3\%. Specifically, the average CV was 4.6\% (maximum 7.5\%) for vitamin~B\textsubscript{12} and 11.0\% (maximum 20.0\%) for iron. These thresholds were used in \citet{cloutier2014aboriginal}. CV values below 16.6\% are considered suitable for unrestricted use, values between 16.6\% and 33.3\% usable with caution, and values above 33.3\% unreliable for publication. 

Point estimate accuracy was assessed using mean absolute error (MAE) and Spearman’s rank correlation. The quality of uncertainty quantification was evaluated using empirical coverage and mean interval length (MIL) of the 90\% credible intervals, as well as the mean interval score (MIS). The MIS, defined as the average Winkler score \citep{winkler1972decision} across all ADM2 regions, captures both sharpness and calibration by combining interval width with a penalty for estimates falling outside the interval. Lower MIS values indicate narrower, well-calibrated intervals.

\subsection{Application to Senagal EHCVM 2021/22 and Nigeria LSS 2018/19}
Following the validation study, we applied the Bayesian SAE models to estimate the prevalence of inadequate intake of folate, iron (Fe), and vitamin B\textsubscript{12} using Senegal’s EHCVM 2021/22 and Nigeria’s LSS 2018/19. Both surveys are designed for ADM1-level analysis. The set of candidate models considered in each country depended on the survey design and the availability of required auxiliary information. Model selection was also guided by findings in the validation study.

Senegal comprises 46 ADM2 departments. The EHCVM follows a two-stage design similar to Rwanda’s EICV, with PPS selection of EAs and systematic random sampling of households, but targeting ADM1-level representativeness. The cluster-level Beta–binomial model considered in this study requires urban–rural composition at the ADM2 level. Although such information is not directly available, urban–rural distributions are available at ADM1 level, from which ADM2-level proportions can be derived (see Appendix~\ref{apx:sen_urban}). Given the compatibility of the survey design and the availability of the required auxiliary inputs, Senegal allowed consideration of the full set of candidate models evaluated in the validation study (cluster-level Beta–binomial, area-level mean-smoothing, and area-level joint smoothing). The model demonstrating the strongest performance in the validation exercise was selected for application.

Nigeria comprises 774 ADM2 areas. Although 741 areas have at least one sampled EA, the large number of administrative units means that many areas contain only a small number of sampled EAs and some have none.
Nigeria presents additional constraints. Like the EHCVM, the LSS is designed for ADM1-level analysis. Compared to the Senegal case, however, aspects of the survey design and available auxiliary information restrict the applicability of the cluster-level Beta–binomial model considered in this paper.
According to the survey technical documentation \citep{NBS2020NLSS}, up-to-date population counts for EAs were not available, with the most recent census conducted in 2006. Consequently, EA selection was not probability proportional to current population size. This has implications for modelling, as households in larger EAs may be under-represented unless survey weights are explicitly incorporated into the analysis. In addition, reliable urban–rural proportions are not available, and up-to-date information is not accessible even at ADM1 level. 
Given the importance of incorporating survey weights in the Nigerian context, we restricted consideration to the area-level models (mean-smoothing and joint smoothing). Among these, the model demonstrating stronger performance in the validation study was selected for application to the LSS data. 

Because reliable ADM2-level direct estimates are generally not available for Senegal and Nigeria, generating stable estimates at this spatial scale is precisely the motivation for applying SAE methods. In the absence of reliable direct ADM2 estimates, we initially assessed the quality of the model-based estimates through visual inspection of posterior means and precision diagnostics based on the coefficient of variation (CV), summarised by the mean CV and the proportion of ADM2 areas exceeding the 16.6\% and 33.3\% thresholds described above.

We then evaluate how well these estimates align with design-based estimates by aggregating ADM2 modelled estimates to the ADM1 level using appropriate population-based geographical weights and comparing them with the corresponding ADM1 direct estimates using MAE, Spearman’s rank correlation, MIL, and empirical coverage.

\section{Results}
This section presents the results of the simulation study using the Rwanda EICV7 survey, followed by the application of the selected SAE models to Senegal and Nigeria. As described previously, the simulation study focused on iron and vitamin B\textsubscript{12}, with folate results presented in Appendix~\ref{apx: folate results}. Folate is included in the application analyses for Senegal and Nigeria, where prevalence levels are higher and greater spatial heterogeneity is observed.

\begin{table*}[t!]
\centering
\begin{tabular}{lcccccccccc}
\toprule
\textbf{Vitamin B}\textsubscript{12} & \multicolumn{5}{c}{Number of sub-sampled EA = 60} & \multicolumn{5}{c}{Number of sub-sampled EA = 30} \\
\cmidrule(lr){2-6} \cmidrule(lr){7-11}
Model & MAE & MIL & Coverage & MIS  & Corr. & MAE & MIL & Coverage & MIS & Corr. \\
\midrule
Direct          & 0.048 & 0.195 & 87.0\% & 0.288 & 0.893 & 0.074 & 0.230 & 74.0\% & 0.651 & 0.804 \\
Area-level &  &  & &  & \\
$\text{ }$Mean-only  & 0.044 & 0.170 & 87.1\% & 0.249 & 0.900& 0.065 & 0.199 & 75.8\% & 0.504 & 0.816 \\
$\text{ }$Joint & 0.043 & 0.191 & 93.0\% & 0.218 & 0.906& 0.060 & 0.259 & 92.3\% & 0.296 & 0.850\\
Custer-level &  &  & &  & \\
$\text{ }$Beta-binomial   & 0.040 & 0.176 & 92.2\% & 0.207 & 0.916 & 0.054 & 0.234 & 92.0\% & 0.276 &  0.863 \\
\bottomrule
\end{tabular}
\caption{Results averaged over 500 experiments for Vitamin B\textsubscript{12}.Mean interval length(MIL), Coverage and  mean interval score(MIS) are all evaluated at the 90\% level.}
\label{tab: validation_results vb12}
\end{table*}

\begin{table*}[t!]
\centering
\begin{tabular}{lcccccccccc}
\toprule
\textbf{Iron} & \multicolumn{5}{c}{Number of sub-sampled EA = 60} & \multicolumn{5}{c}{Number of sub-sampled EA = 30} \\
\cmidrule(lr){2-6} \cmidrule(lr){7-11}
Model & MAE & MIL & Coverage & MIS  & Corr. & MAE & MIL & Coverage & MIS & Corr. \\
\midrule
Direct          & 0.038 & 0.179 & 85.6\% & 0.228 & 0.834 & 0.057 & 0.174 & 72.3\% & 0.337 & 0.709\\
Area-level &  &  & &  & \\
$\text{ }$Mean-only  & 0.032 & 0.147 & 85.6\% & 0.189 & 0.843 & 0.049 & 0.146 & 73.9\% & 0.269 & 0.712\\
$\text{ }$Joint & 0.032 & 0.158 & 88.5\% & 0.170 & 0.833 & 0.045 & 0.159 & 83.2\% & 0.200 & 0.736\\
Custer-level &  &  & &  & \\
$\text{ }$Beta-binomial   & 0.031 & 0.159 & 92.1\% & 0.159 & 0.853 & 0.040 & 0.174 & 91.8\% & 0.176 & 0.756\\
\bottomrule
\end{tabular}
\caption{Results averaged over 500 experiments for Iron. Mean interval length(MIL), Coverage and  mean interval score(MIS) are all evaluated at the 90\% level. The correlation $\rho$ shown is Spearman's rank correlation.}
\label{tab: validation_results fe}
\end{table*}

\subsection{Rwanda simulation study}\label{sec: experiment results}
Across all scenarios, the modelled estimates improved upon the direct estimates obtained from the sub-sampled surveys (Table~\ref{tab: validation_results vb12} and Table~\ref{tab: validation_results fe}). These improvements were most evident in the data-sparse setting (30 EAs per ADM1), where both area-level and cluster-level models substantially reduced MAE and MIS, indicating gains in accuracy and better-calibrated uncertainty. Under the less sparse design (60 EAs per ADM1), improvements in MAE were more modest; however, consistent gains in coverage and MIS were still observed, particularly for the area-level joint smoothing model and the cluster-level Beta–binomial model.

For vitamin~B\textsubscript{12} in the most data-sparse setting, the joint smoothing model produced wider intervals than the direct estimates (an increase of 12.6\% in MIL), but achieved a substantially lower MIS (a reduction of 54.5\%). This reflects the fact that the apparent advantage of the direct estimates arose from overly narrow intervals that were often far from the true values.

Among the models considered, the cluster-level Beta-binomial model achieved the lowest MAE and MIS, with high coverage and the strongest correlation with the gold-standard design-based estimates. This approach performs well when cluster-level information for the entire population is available, although such data are not always available in all countries, and the model does not explicitly incorporate survey design weights. The area-level models, by contrast, are design-consistent and do not require cluster-level information. Between the two area-level approaches, the joint smoothing model showed stronger performance than the mean smoothing model across all scenarios with improvements particularly evident in the data-sparse setting, where sampling variance was higher. For example, relative to the mean smoothing model, the joint smoothing model reduced the MIS by 41.3\% for vitamin~B\textsubscript{12} and 25.7\% for iron. The mean smoothing model tended to exhibit coverage below the nominal level, a finding consistent with \citet{gao2023spatial}. For iron, which displayed higher overall variability, the joint smoothing model also yielded coverage metrics lower than the nominal level, but still performed better than the alternative mean smoothing model. 

Overall, the cluster-level Beta–binomial model demonstrated the strongest performance  in this evaluation setting, while the area-level joint smoothing model provided the most reliable alternative among the models that account for survey design. Guided by these findings, alongside survey design considerations, the cluster-level Beta–binomial model was selected for application to Senegal, whereas for Nigeria, where survey design considerations favour incorporation of survey weights, the area-level joint smoothing model was applied.

\subsection{Senegal (EHCVM 2021/22)}
\begin{figure*}[th]
    \centering
    \includegraphics[width=0.99\linewidth]{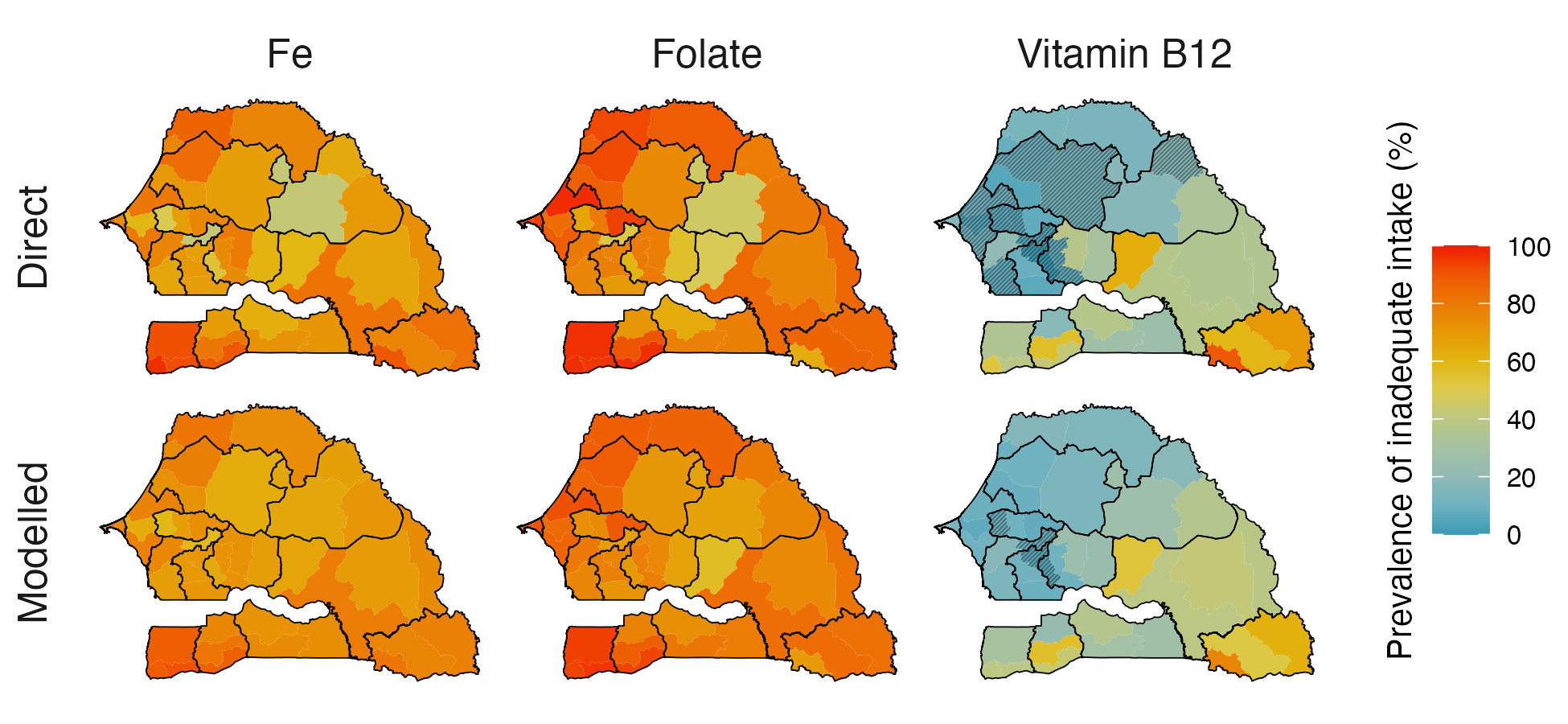}
    \caption{Estimated prevalence of inadequate micronutrient intake in Senegal for iron (Fe), folate and vitamin B\textsubscript{12}. The top row shows ADM2-level direct survey estimates, and the bottom row shows the corresponding modelled estimates obtained using the cluster-level Beta–binomial model.  Values are constant within each ADM2 polygon and represent area-level estimates rather than a continuously varying spatial surface. Regions where the direct coefficient of variation (CV) exceeds 33.3\% are indicated with hatching.}
    \label{fig:senegal prevalance map}
\end{figure*}
\begin{figure*}[th]
    \centering
    \includegraphics[width=0.99\linewidth]{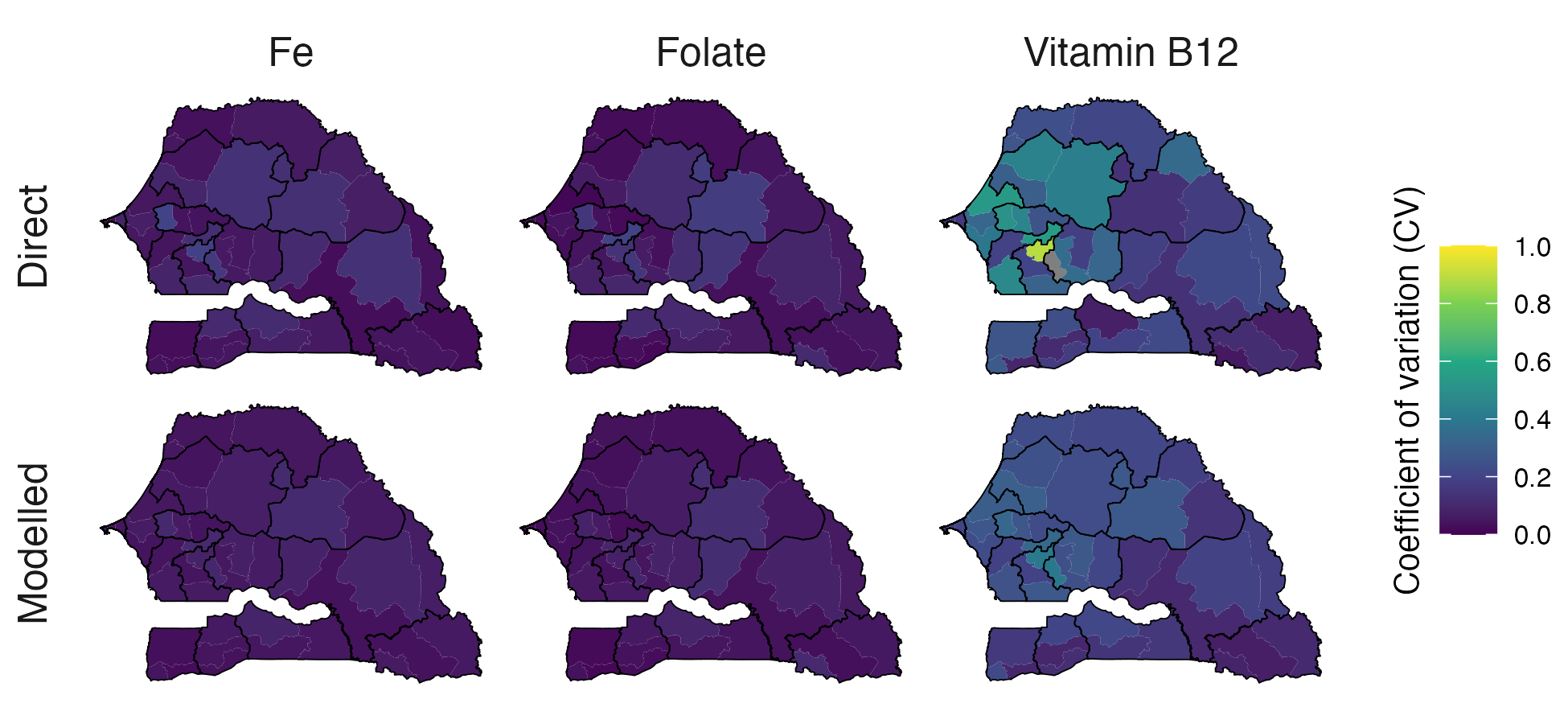}
    \caption{Coefficient of variation (CV) for ADM2-level estimates of inadequate intake of iron (Fe), folate and vitamin B\textsubscript{12} in Senegal. The top row shows CVs from the direct survey estimator, and the bottom row shows the corresponding CVs from the cluster-level Beta–binomial model. Values are constant within each ADM2 polygon and represent area-level estimates rather than a continuously varying spatial surface. Areas with no sampled clusters are shown in grey.}
    \label{fig:senegal cv map}
\end{figure*}

\begin{table*}[th]
\centering
\begin{tabular}{lcccccccc}
\hline
\multirow{2}{*}{Nutrient} & \multicolumn{2}{c}{MIL} & \multicolumn{2}{c}{Mean CV} & \multicolumn{2}{c}{CV \textgreater 16.6\%} & \multicolumn{2}{c}{CV \textgreater 33.3\%} \\ \cmidrule{2-9} 
 & Direct & Modelled & Direct & Modelled & Direct & Modelled & Direct & Modelled \\ \hline
Fe & 0.158 & 0.135 & 0.0718 & 0.0576 & 2 & 0 & 0 & 0 \\
Folate & 0.150 & 0.128 & 0.0640 & 0.0514 & 2 & 0  & 0 & 0 \\
Vitamin B12 & 0.157 & 0.138 & 0.287 & 0.212 & 35 & 32 & 14 & 5 \\ \hline
\end{tabular}
\caption{Summary of ADM2-level uncertainty measures of the direct and modelled (cluster-level Beta–binomial model) estimates for Senegal. MIL denotes the mean interval length of the 90\% credible intervals. Mean CV refers to the average coefficient of variation across ADM2 areas, and the final two columns report the proportion of ADM2 units with CV exceeding 16.6\% and 33.3\%, respectively.}
\label{tab: uncertainty senagal}
\end{table*}

\begin{figure*}[th]
    \centering
    \includegraphics[width=0.99\linewidth]{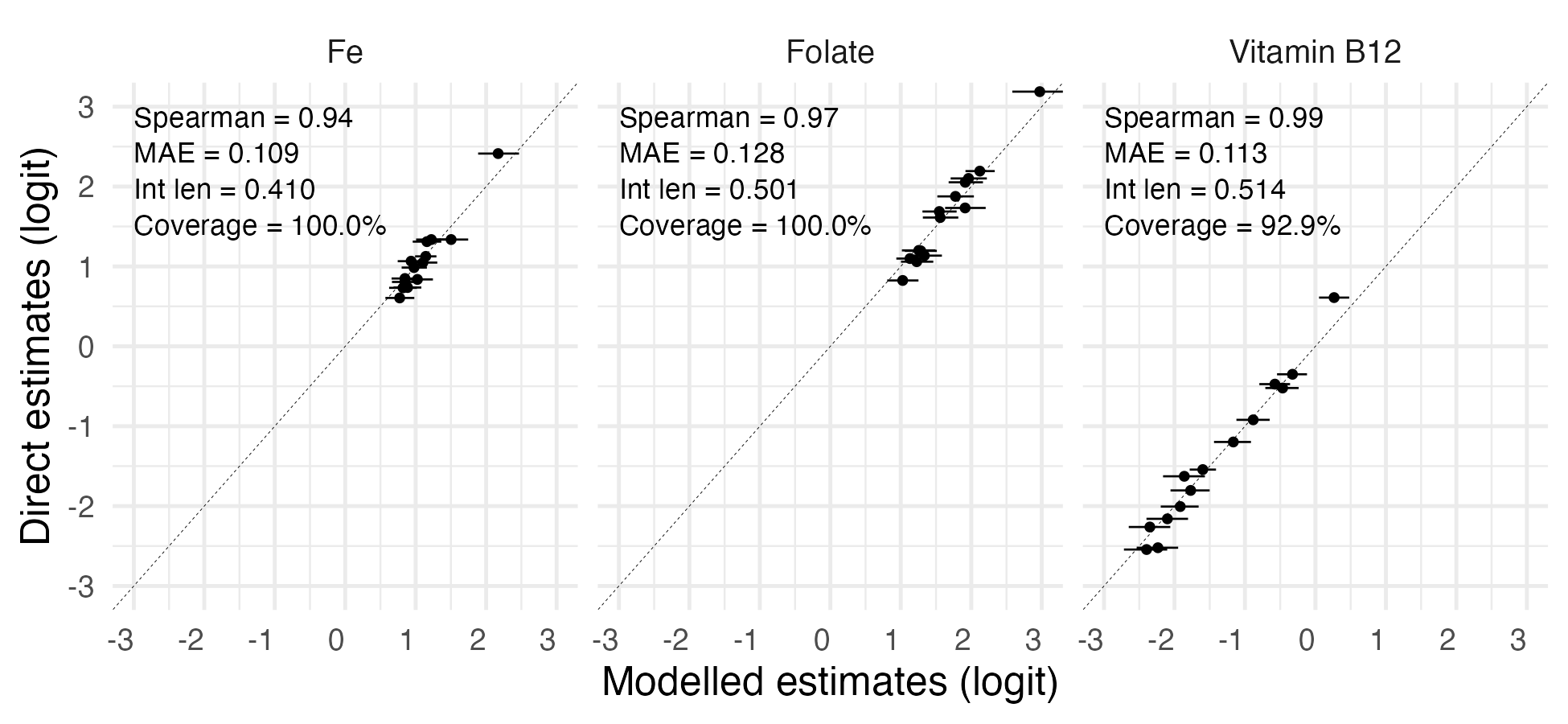}
    \caption{Comparison of ADM1-level direct estimates for Senegal, against modelled estimates obtained by aggregating the ADM2-level estimates from the cluster-level Beta–binomial model to the ADM1 level using population-based geographical weights. Both estimates are in logit scale. Each panel corresponds to a different micronutrient (iron, folate, and vitamin B\textsubscript{12}). Points represent ADM1-level estimates, with 90\% credible intervals shown for the model-based estimates. Reported statistics include the Spearman rank correlation, mean absolute error (MAE), mean interval length (Int len) of the 90\% credible intervals, and empirical coverage.}
    \label{fig:senegal adm1 scatter}
\end{figure*}
Using the selected cluster-level Beta–binomial model, we produced ADM2-level estimates of the prevalence of micronutrient inadequacy for Senegal.
The results for Senegal are summarised in Figures \ref{fig:senegal prevalance map}, \ref{fig:senegal cv map}, \ref{fig:senegal adm1 scatter} and Table~\ref{tab: uncertainty senagal}. The ADM2-level prevalence maps in Figures~\ref{fig:senegal prevalance map} highlight spatial patterns. Iron and folate inadequacy are generally high across the country, with particularly elevated levels in the south-west and north-west, and an even stronger concentration of high folate inadequacy in the south-west. In contrast, vitamin B\textsubscript{12} exhibits a markedly different spatial structure, with lower inadequacy in the north-west and relatively higher levels in the south-east. As shown in Figure \ref{fig:senegal cv map}, the CVs of the direct estimates are generally low for iron and folate but substantially higher for vitamin B\textsubscript{12}, reflecting its lower prevalence. Across all three nutrients, the modelled estimates based on the cluster-level Beta–binomial model display smoother spatial patterns and noticeably reduced uncertainty.

We first examine the uncertainty of the ADM2-level modelled estimates (Table \ref{tab: uncertainty senagal}) and then assess alignment with design-based estimates by aggregating ADM2-level modelled estimates to the ADM1 level and comparing them with the corresponding ADM1 direct estimates (Figure \ref{fig:senegal adm1 scatter}). At the ADM2 level, the model substantially reduces uncertainty across all three micronutrients, with decreases in MIL and mean CV. For iron and folate, a small number (two) of ADM2 areas exceed the 16.6\% CV threshold under direct estimation, but none do so under the model. For vitamin B\textsubscript{12}, the proportion of ADM2 units with CV above 16.6\% decreases from 77.8\% to 71.1\%, while the proportion above 33.3\% drops from 28.9\% to 8.9\%. Notably, ADM2 units with the highest vitamin B\textsubscript{12} inadequacy are estimated with acceptable precision, which is important for operational use. At the ADM1 level, aggregated modelled estimates align closely with direct estimates, showing high Spearman correlations, low MAE, reasonable interval lengths, and good empirical coverage.

\begin{table*}[th]
\centering
\begin{tabular}{lcccccccc}
\hline
\multirow{2}{*}{Nutrient} & \multicolumn{2}{c}{MIL} & \multicolumn{2}{c}{Mean CV} & \multicolumn{2}{c}{CV \textgreater 16.6\%} & \multicolumn{2}{c}{CV \textgreater 33.3\%} \\ \cmidrule{2-9} 
 & Direct & Modelled & Direct & Modelled & Direct & Modelled & Direct & Modelled \\ \hline
Fe & 0.182 & 0.236 & 0.124 & 0.146 & 206 & 265 & 55 & 3 \\
Folate & 0.188 & 0.231 & 0.121 & 0.138 & 189 & 220 & 42 & 0 \\
Vitamin B12 & 0.156 & 0.202 & 0.130 & 0.150 & 205 & 313 & 68 & 6 \\ \hline
\end{tabular}
\caption{Summary of ADM2-level uncertainty measures of the direct and modelled (area-level joint smoothing model) estimates for Nigeria. MIL denotes the mean interval length of the 90\% credible intervals. Mean CV refers to the average coefficient of variation across ADM2 areas, and the final two columns report the proportion of ADM2 units with CV exceeding 16.6\% and 33.3\%, respectively.}
\label{tab: uncertainty nigeria}
\end{table*}

\begin{figure*}[th]
    \centering
    \includegraphics[width=0.99\linewidth]{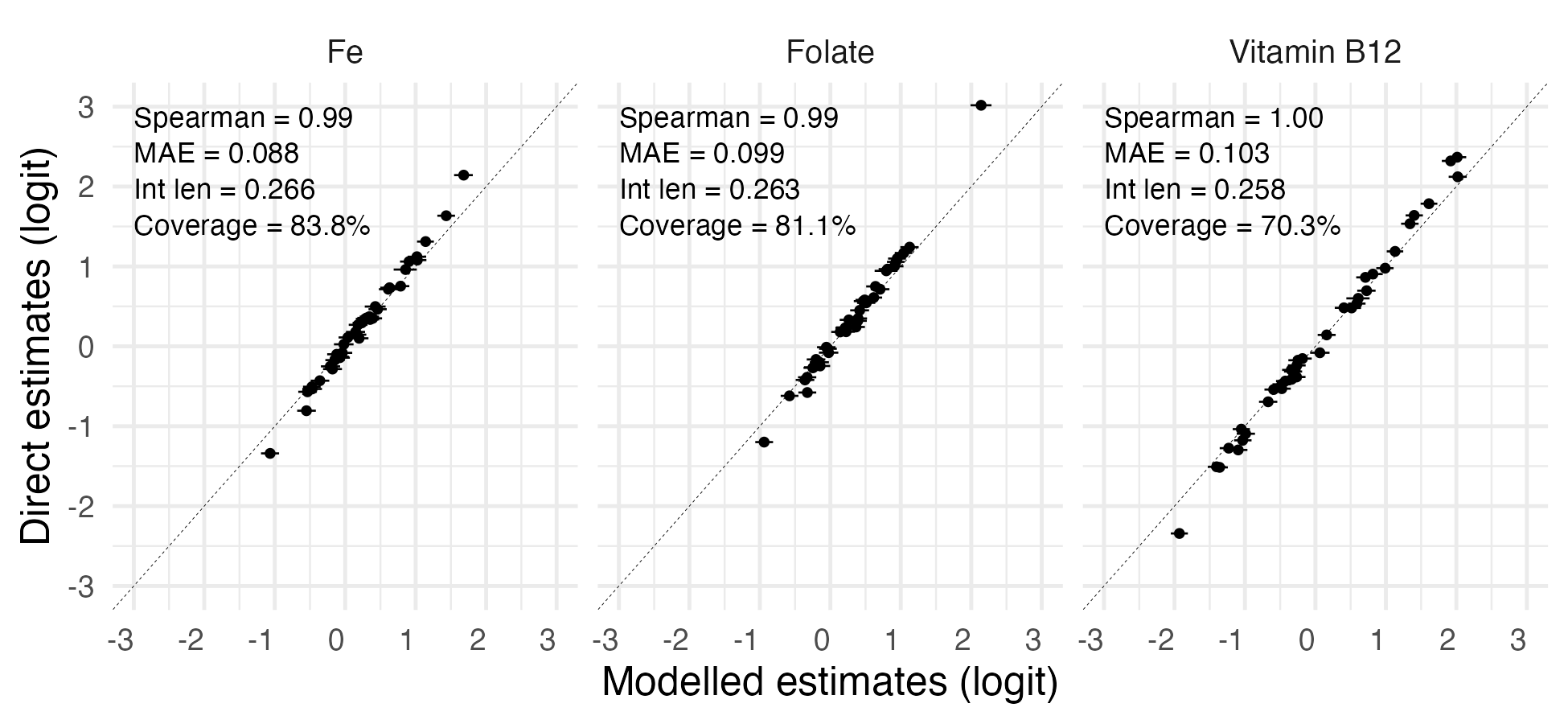}
    \caption{Comparison of ADM1-level direct estimates for Nigeria, against modelled estimates obtained by aggregating the ADM2-level estimates from the area-level joint-smoothing model to the ADM1 level using population-based geographical weights. Both estimates are in logit scale. Each panel corresponds to a different micronutrient (iron, folate, and vitamin B\textsubscript{12}). Points represent ADM1-level estimates, with 90\% credible intervals shown for the model-based estimates. Reported statistics include the Spearman rank correlation, mean absolute error (MAE), mean interval length (Int len) of the 90\% credible intervals, and empirical coverage.}
    \label{fig:nigeria adm1 scatter}
\end{figure*}

\subsection{Nigeria (LSS 2018/19)}
\begin{figure*}[th]
    \includegraphics[width=0.99\linewidth]{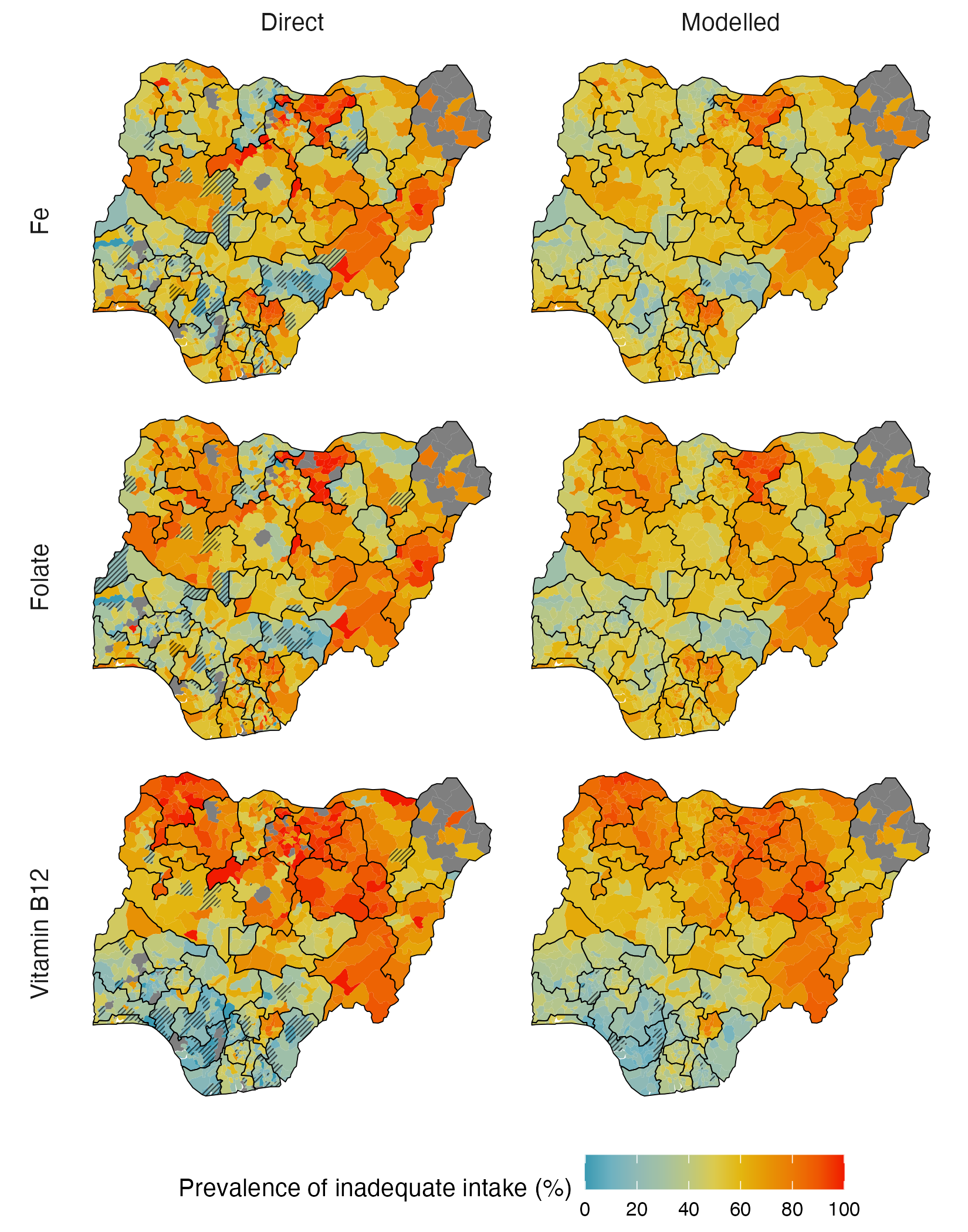}
        \caption{Estimated prevalence of inadequate micronutrient intake in Nigeria for iron (Fe), folate and vitamin B\textsubscript{12}. The left panels show ADM2-level direct survey estimates, and the right panels show the corresponding modelled estimates obtained using the area-level joint smoothing model. Values are constant within each ADM2 polygon and represent area-level estimates rather than a continuously varying spatial surface. Regions where the direct coefficient of variation (CV) exceeds 33.3\% are indicated with hatching. Modelled estimates are not shown for parts of Borno due to insufficient sample.}
    \label{fig:nigeria prevalence map} 
\end{figure*}

\begin{figure*}[th]
    \includegraphics[width=0.99\linewidth]{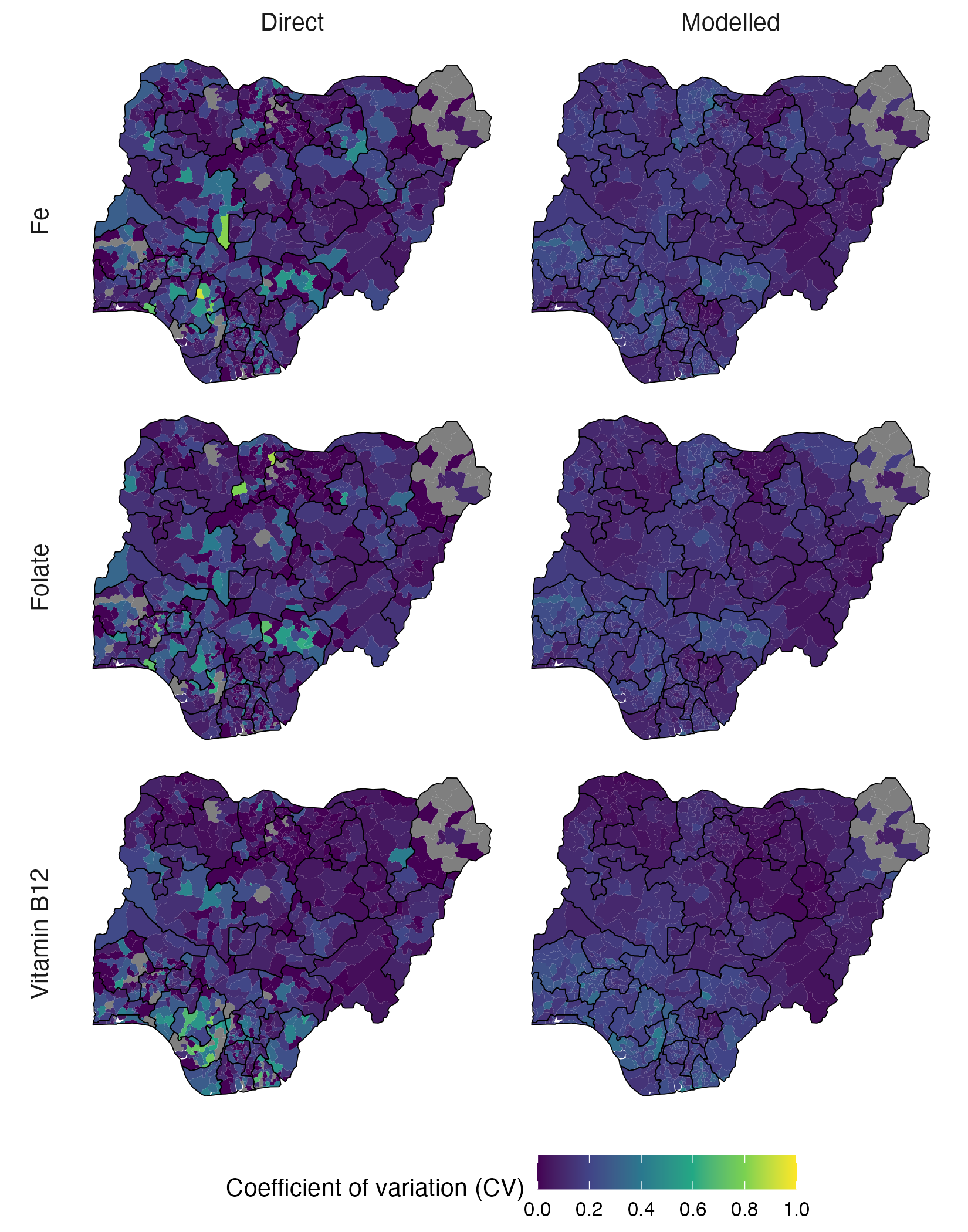}
        \caption{Vitamin B\textsubscript{12} inadequate intake in Nigeria. The top panel shows ADM2-level prevalence estimates from the direct survey estimator (left) and the area-level joint smoothing model (right). The bottom panel presents the corresponding coefficients of variation (CV). Values are constant within each ADM2 polygon and represent area-level estimates rather than a continuously varying spatial surface. Areas with no valid direct variance estimates are shown in grey, and in the top panel, regions where the CV exceeds 33.3\% are indicated with hatching. Modelled estimates are not shown for parts of Borno due to insufficient sample.}
    \label{fig:nigeria cv map}
\end{figure*}
ADM2-level estimates of the prevalence of micronutrient inadequacy for Nigeria were obtained using the selected area-level joint smoothing model.
The results for Nigeria are summarised in Table~\ref{tab: uncertainty nigeria} and Figures~\ref{fig:nigeria adm1 scatter}, \ref{fig:nigeria prevalence map}, \ref{fig:nigeria cv map}. The ADM2-level prevalence maps in Figure~\ref{fig:nigeria prevalence map} show spatial heterogeneity across all three micronutrients. Iron and folate inadequacy are widespread, with some spatial heterogeneity where pockets of low inadequacy are interspersed with areas of higher inadequacy. In contrast, vitamin B\textsubscript{12} inadequacy displays clearer spatial structure, with higher inadequacy observed in Northern and Eastern Nigeria, compared with the rest of the country. Compared with Senegal, the direct estimates in Nigeria exhibit a more fragmented spatial structure, reflecting the much larger number of ADM2 units and the limited number of sampled EAs within many ADM2 areas.

In Figure~\ref{fig:nigeria cv map}, we can observe high uncertainty under the direct estimator across Nigeria, particularly for vitamin B\textsubscript{12}, with many ADM2 units exhibiting very high CVs. Iron and folate also show widespread moderate-to-high uncertainty at the ADM2 level. The modelled estimates obtained from the area-level joint smoothing model produce noticeably smoother spatial patterns and reduce extreme values, although uncertainty remains non-negligible in many areas. This reflects the challenging Nigerian survey setting.

Table~\ref{tab: uncertainty nigeria} summarises ADM2-level uncertainty measures for the direct and modelled estimates. Unlike the Senegal case, the modelled estimates are associated with larger mean interval lengths and higher mean CVs across all three micronutrients. In addition, the number of ADM2 units exceeding the 16.6\% CV threshold increases under the modelled estimates. 
Importantly, however, the model delivers a substantial reduction in extreme uncertainty. For iron, folate, and vitamin B\textsubscript{12}, the number of ADM2 units with CV exceeding 33.3\% drops substantially under the modelled estimates, from 55 to 3, 42 to 0, and 68 to 6, respectively.

As with Senegal, direct and modelled estimates cannot be compared directly at the ADM2 level because the LSS is designed for ADM1 analysis. The aggregated modelled estimates show strong agreement with the ADM1 direct estimates across all three nutrients (Figure~\ref{fig:nigeria adm1 scatter}), with high Spearman’s rank correlations, indicating that the joint smoothing model preserves fidelity to the direct, design-based estimates at the scale for which the survey is designed. However, empirical coverage of the aggregated 90\% credible intervals falls below the nominal level, particularly for vitamin B\textsubscript{12}. We also observe some shrinkage towards the overall mean. This limitation will be discussed in the following section.

\section{Discussion}
This study demonstrates that Bayesian SAE methods can be used to generate estimates of the prevalance of inadequate micronutrient intake using routinely collected HCES data at a finer spatial scale. Such information can strengthen the evidence base for more informed prioritisation and targeting of interventions by highlighting geographic variation in risk. Using ADM2 representative data, we evaluated two Bayesian SAE model classes, the cluster-level Beta–binomial model and area-level models, under realistic sampling conditions. These models were then applied to two countries (Senegal \& Nigeria) to illustrate their practical utility in real-world settings where the survey is designed only for ADM1 analysis. 

To ensure the responsible development of these models, we evaluated both cluster-level and area-level Bayesian SAE models using data from the Rwanda EICV7 survey, which is designed to support ADM2-level inference. The simulation study was conducted under realistic, data-constrained scenarios with two sampling intensities and focused on three micronutrients: iron, folate, and vitamin B12. For both model classes across all scenarios, the Bayesian SAE approaches consistently provided substantial improvements over direct estimates, particularly in terms of reducing estimation error and providing well-calibrated credible intervals. Given the instability of direct ADM2 estimates arising from small sample sizes obtained from surveys not designed for inference at this level \citep{jiang2020robust}, these Bayesian SAE models offer a more reliable way to characterise subnational patterns at the ADM2 level.

The cluster-level Beta–binomial model generally achieved higher accuracy and better uncertainty calibration than the area-level models. However, as a unit-level approach, it requires careful handling of complex survey designs, and incorporating survey weights is not always straightforward \citep{gelman2007struggles, parker2023comprehensive}.  In settings where selection probabilities vary with cluster size, ignoring weights may introduce bias. By contrast, the area-level joint smoothing model, which smooths both prevalence estimates and their sampling variances, remains compatible with survey design and still offers clear improvements over direct estimation.  Accordingly, we applied the cluster-level model in Senegal and the area-level model in Nigeria, reflecting differences in survey design and the role of weights.

These results are not intended to establish a universal ranking of models, but to guide model choice conditional on survey design. The use of the simulation study to inform model selection is based on the assumption that survey design and the amount of information available for ADM2 estimation are important determinants of model behaviour. The two simulation scenarios therefore represent different levels of ADM2 sampling sparsity relevant to the application surveys, and the results should be interpreted accordingly.
Cluster-level models are more suitable when cluster-level information is available and sampling designs do not require explicit weighting, whereas area-level models are preferable when weights must be incorporated or auxiliary data are limited. The applications to Senegal and Nigeria reflect these considerations.

Our application in Senegal demonstrated that the cluster-level Beta–binomial model produced ADM2 estimates with substantially reduced uncertainty compared with direct estimates, while preserving strong consistency with ADM1 design-based estimates after aggregation. Estimates generated using this approach become usable for nutrition programme and policy decision makers, providing higher resolution insights for the geographical targeting of programmes implemented at this level where this may otherwise not be feasible. 

Nigeria represents a challenging application context due to its large number of ADM2 units, limited sample sizes in many of these units, and the increased importance of appropriately accounting for survey weights given the survey sampling design. Despite these constraints, the area-level joint smoothing models delivered important practical improvements compared with direct estimation by substantially reducing the number of ADM2 areas with extreme coefficients of variation (above $33\%$). This reduction is particularly relevant for operational use, as it substantially limits the number of areas for which ADM2 estimates would be considered too unreliable for policy interpretation. In addition, the models reduced small-area noise, and maintained agreement with ADM1 direct estimates when aggregated. However, in this context, the area-level joint smoothing model did not uniformly reduce average uncertainty measures, and exhibited some degree of shrinkage toward the overall mean which may mask true local extremes. This could potentially be problematic when applying this model to to other highly data-constrained settings similar to Nigeria, especially if pockets of high vulnerability within otherwise low- or moderate-burden regions are hidden, obscuring the sub-populations with the greatest needs. For this reason, results from area-level joint smoothing models should be interpreted with care, particularly in highly data-sparse settings. In addition to examining point estimates and standard uncertainty measures, it is also important to consider the potential effects of model-based smoothing on bias and interval coverage. Alternative modelling strategies, such as incorporating additional covariates or adopting unit-level approaches that appropriately account for survey weights , may be preferable. Recent methodological developments have proposed approaches for using survey weights in cluster-level binomial models \citep{chen2014use} or unit-level model in general \citep{parker2023comprehensive} and could be considered for application in settings similar to the Nigeria context.

Taken together, these findings indicate that even under highly constrained survey designs, area-level Bayesian SAE can yield usable subnational estimates, provided that the implications of smoothing and shrinkage are carefully assessed.

\subsection{Policy implications}
Globally, micronutrient inadequacy is unevenly distributed across populations \citep{passarelli2024global}. Government nutrition strategies are often structured around national programmes that aim to increase micronutrient intake through food systems interventions (e.g., food fortification) or health systems interventions (e.g., supplementation) \citep{WHO2018GlobalNutritionPolicyReview}. While effective when delivered as intended, vulnerable groups may have persistent unmet need, as evidenced by geographic and sociodemographic disparities in vulnerability to inadequate micronutrient intake and intervention coverage \citep{tang2023evaluating, tang2025potential}. In areas where elevated dietary inadequacy coincides with low programme coverage, unmet need becomes geographically concentrated requiring targeted prioritisation or further intervention.
One approach to address remaining micronutrient need is to integrate nutrition policies aimed at increasing micronutrient intake into existing social assistance programmes that reach vulnerable populations (e.g., school-based programmes, cash and voucher transfers, food transfers), leveraging delivery platforms that can support nutrition-sensitive social protection \citep{ruel2013nutrition}. Because many of these programmes operate at the ADM2 level, effective decisions about where and how to integrate micronutrient interventions depend on having evidence on dietary inadequacy at the same spatial scale. The Bayesian SAE methods described provide ADM2-level evidence on dietary inadequacy that can inform which existing programmes and locations should be prioritised for integrating micronutrient interventions.

However, it is important to emphasise that ADM2-level estimates modelled using Bayesian SAE frameworks should not be viewed as a replacement for high-resolution primary data collection of dietary data at the same spatial scale. While the SAE models described can stabilise estimates and quantify uncertainty in contexts where granular data are not available, they remain conditional on modelling assumptions and ultimately cannot fully recover fine-scale variation that is not captured by the original underlying data sample. For this reason, continued investment in high-resolution primary data collection remains essential for the generation of evidence which can guide programme design and evaluation. However, in contexts where these high-resolution data are not currently available, ADM2 estimates derived from SAE models offer an alternative that can be used to inform nutrition programme and policy decision making.

\subsection{Study limitations}
This study had some limitations. 
First, the models presented in this study rely primarily on spatial neighbourhood structure and limited auxiliary information, using only urban or rural stratification for cluster-level models. Additional covariates, such as poverty and market access, which are known drivers of dietary inadequacy \cite{tang2022modeling}, were not incorporated, as comparable and reliable data were not consistently available across the three country contexts, particularly at the spatial resolution required for modelling. While this simplifies the modelling framework, it may contribute to over-smoothing. As such, fine-scale variation that is driven by non-spatial factors may be obscured by the modelled estimates. Further research is therefore needed to extend and validate the Bayesian SAE models to incorporate additional covariates. In settings where conventional socio-economic data are unavailable or difficult to access, satellite-derived covariates may provide a scalable alternative source of auxiliary information to better capture spatial heterogeneity \cite{newhouse2025small}. 

In addition, dietary inadequacy was estimated using apparent intake derived from HCES data, which is a household-level proxy that requires several assumptions, including proportional intra-household food distribution according to household member energy requirements. As a result, the outcome variable used in the SAE models reflects intake proxies rather than more precise individual intake, where inadequacy arising from individual-level variation in diets is unaccounted. 

\subsection{Conclusion}
In summary, this study demonstrates how Bayesian SAE frameworks can be applied to routinely collected population surveys such as HCES to generate actionable ADM2 estimates of micronutrient inadequacy, informing nutrition programme planning at the subnational level. While not a substitute for high-resolution primary data, these methods offer a precision public health approach that can provide decision-makers with the evidence required to prioritise populations with the greatest need, supporting more equitable resource allocation.

\backmatter
\bmhead{Acknowledgements}
This project is supported by Engineering and Physical Sciences Research Council (EPSRC) under grant EP/V002910/2 and by the Gates Foundation (INV-037325) through the Modelling and mapping the risk of Inaequate Micronutrient Intake (MIMI) project. The authors thank Dr Emily Nightingale for her critical review of the manuscript and for her valuable comments.
\bmhead{Author contributions}
SI, MO, SF and KT were responsible for the overall study design. SI was responsible for data analysis across all countries, interpretation of results, and manuscript writing. MO was responsible for data preprocessing across all countries, contributed to data analysis and interpretation of results, and was responsible for manuscript writing. ZC contributed to the design, analysis, and interpretation of the validation study using Rwanda data. UA contributed to data preprocessing in Senegal and Rwanda. EB contributed to data preprocessing in Senegal and Rwanda. GB contributed to data preprocessing in Nigeria. DH contributed to data preprocessing in Senegal. DP contributed to the conception of the study, and interpretation of results. FK contributed to the conception of the study and made substantial revisions to the manuscript. SF contributed to the conception of the study, interpretation of results and provided supervision for the analysis. KT provided overall supervision for the study, contributed to the conception of the study, interpretation of results and was a major contributor in writing the manuscript. All authors read and approved the final manuscript.
\bmhead{Availability of data and materials}
All code used for data preprocessing, model implementation, and analysis is openly available at
\url{https://github.com/MIMI-wfp/mimi_spatial_smoothing}. 

The Rwanda EICV7 dataset is publicly available subject to registration and approval from the National Institute of Statistics Rwanda at \url{https://statistics.gov.rw/datasource/integrated-household-living-conditions-survey-eicv7}

The Senegal EHCVM dataset is publicly available subject to registration and approval on the World Bank's Microdata Library at \url{https://microdata.worldbank.org/catalog/6278/get-microdata}. 

The Nigeria LSS dataset is publicly available subject to registration and approval on the World Bank's Microdata Library at \url{https://microdata.worldbank.org/catalog/3827/get-microdata}.

\section*{Declarations}
\paragraph{Competing interests}
The authors declare that they have no competing interests.
\paragraph{Ethics approval and consent to participate}
Not applicable
\paragraph{Consent for publication}
Not applicable

\clearpage
\begin{appendices}
\section{}
\subsection{Estimating household apparent micronutrient intake}\label{sec: Indicator construction}
Household apparent micronutrient intake was estimated using the nutrient supply model which consists of three key parameters that represent: (i) household consumption of food-items (Section \ref{sec: food-item composition}), (ii) the nutrient composition of these food items (Section \ref{sec: nutrients composition}), and (iii) the demographic composition of households (Section \ref{sec: hh demographic}) \citep{fiedler2012hces}. These parameters were combined to estimate apparent household micronutrient intake of iron, folate, and vitamin B12 (Section \ref{sec: apparent hh mi}).
\subsubsection{Food-item consumption quantities}\label{sec: food-item composition}
Food-item consumption quantities were derived from the household food consumption module of each survey. In all three instances, consumption was captured over a 7-day period. In cases where consumption of food-items was reported using non-standard units such as ``piece'' or ``bunch'', quantities were converted to metric units using conversion factors. Where relevant, quantities were also adjusted using edible-portion factors derived from food composition tables to account for inedible parts of certain food items (e.g. the skin of a banana). Reported 7-day consumption of each food item for each household was then divided by seven to obtain daily household consumption ($Q_{ih}$) in grams per day (g/day) for each food item $i$ and household $h$.
\subsubsection{Nutrient composition of food-items}\label{sec: nutrients composition}
Food-items from each survey were matched to corresponding food-items in the relevant food composition tables detailed in section \ref{sec: Data sources}. For each food-item $i$, nutrient composition values ($C_i$) were derived (e.g. mcg of folate per 100 grams of rice).
\subsubsection{Household demographics}\label{sec: hh demographic}
To enable meaningful comparison between households with different sizes and demographic structures, household nutrient intake values were standardised and estimates were presented as nutrient intake per Adult Female Equivalent ($AFE_h$) for each household $h$ \citep{weisell2012AME, kalimbira2025AFE}.

The $AFE_h$ unit is a measure that expresses the energy and nutrient requirements of all individuals in a household relative to those of one non-pregnant, non-lactating adult female aged 18-29 years old with moderate physical activity, denoted as $R_{AFE}$ \cite{EnergyRequirements}. Requirements for other individuals indexed by $j$ are expressed as $R_j$ and are estimated using their age and sex as reported in the household roster.

To calculate $AFE_h$, the sum of each individuals energy requirements $R_j$ relative to the reference individual $R_{AFE}$ is taken for $m_h$ individuals residing in household $h$, and is shown by
\begin{equation}\label{eq: adult female equivalent}
    AFE_h = \sum_{j=1}^{m_h} \frac{R_j}{R_{AFE}}.
\end{equation}

\subsubsection{Apparent micronutrient intake}\label{sec: apparent hh mi}
The parameters detailed above were then combined to model the apparent household micronutrient intake $I_h$ as
\begin{equation}\label{eq: household apparent intake}
    I_h = \frac{\sum_{i=1}^{n} Q_{ih}C_{i}}{AFE_h}
\end{equation}
where $n$ denotes the number of food-items.

\subsection{Sampling variance}\label{apx: sampling variance}

For area-level SAE models, the sampling variance of the direct prevalence estimator $\hat{p}_\ell$, denoted by $\hat{V}_\ell$, is required for all areas $\ell = 1,\ldots,L$. For the Rwanda simulation study and the Senegal case study, we estimated $\hat{V}_\ell$ using the \texttt{survey} package in \textsf{R} \cite{lumley2024survey}, which implements Taylor series linearisation variance estimators for complex survey designs. Practical implementation details are provided in \cite{lumley2004analysis}, with theoretical foundations discussed in \cite{binder1983variances,wolter2007introduction}. In this setting, the degrees of freedom $d_\ell$ are taken to be the number of sampled clusters (EAs) in area $\ell$ minus one.

A complication arises for areas in which only a single cluster is sampled, for which the design-based variance estimator is undefined or degenerates to zero. In such cases, we augment the data with a \emph{phantom cluster} \cite{dong2025principledworkflowprevalencemapping}. The phantom cluster is assigned a prevalence equal to that of the corresponding ADM1 area containing the ADM2 area, and its total survey weight is set to the average cluster-level weight observed within that ADM1 area. This strategy stabilises the variance estimation while preserving consistency with higher-level survey information, and follows an approach similar to that described in \cite{dong2025principledworkflowprevalencemapping}.

For the Nigeria case study, a substantial proportion of ADM2 areas were affected by this issue. Rather than introducing a large number of phantom clusters, we instead approximated the sampling variance using a binomial-type expression based on an effective sample size, given by

\begin{equation}
\hat{V}_\ell = \frac{\hat{p}_\ell(1-\hat{p}_\ell)}{n_\ell^*}, \notag
\end{equation}
where $n_\ell^* = \frac{1}{\sum_{h \in \mathcal{S}_\ell} \tilde{w}_h^2}$ denotes the effective sample size in area $\ell$, and $\tilde{w}_h$ are survey weights normalised to sum to one within area $\ell$. The corresponding degrees of freedom are taken to be $d_\ell = n_\ell^* - 1$.

Note that, for sense checking at the ADM1 level, comparisons between aggregated model-based estimates and direct survey estimates are conducted on the logit scale. Let $\hat{p}_j$ denote the direct ADM1 prevalence estimate with sampling variance $\hat{V}_j$. The sampling variance of the logit-transformed estimate is approximated using a first-order delta method and given by ${\hat{V}_j}/ \left(\hat{p}_j^2(1-\hat{p}_j)^2\right)$.

\subsection{Sub-sampling procedure of Rwanda EICV survey}\label{apx: subsampling procedure}
The Rwanda simulation study relies on a household survey originally designed to support design-based inference at the ADM2 level under a two-stage cluster sampling design. To evaluate model performance when survey data support inference only at the ADM1 level, we construct sub-samples that mimic a household survey designed for ADM1 analysis, while preserving the original clustered sampling structure and sampling frame.

Sub-sampling is conducted at the EA level using a stratified, probability-weighted sampling scheme. Within each ADM1 unit, EAs are first stratified by urban–rural status, and a fixed total number of EAs is allocated to the two strata in proportion to the number of urban and rural EAs in that ADM1, based on external administrative information. Within each ADM1 $\times$ urban–rural stratum, EAs are sampled without replacement with probabilities proportional to the original survey weights, which serve as proxies for the realised first-stage inclusion probabilities under the ADM2 survey design. All households within selected EAs are retained, yielding a sub-sample that preserves clustering, ADM1-level representativeness, and urban–rural composition, while approximating a household survey designed for ADM1-level inference.

\subsection{Additional results for Rwanda experiments}\label{apx: folate results}
The prevalence of the inadequate folate intake is in general low in Rwanda.
Although there may be less interest in obtaining modelled estimates from a policy perspective, we conducted validation study with folate in addition to Iron and Vitamin $\text{B}_{12}$. The result is presented in Table~\ref{tab: Rwanda folate result}. 
\begin{table}[t]
    \centering
    \resizebox{\linewidth}{!}{
    \begin{tabular}{lccccc}
    \hline
    Model & MAE & MIL & Coverage & MIS & Corr.\\
    \hline
    Direct & 0.019 & 0.073 & 0.882 & 0.155 & 0.684\\
    Area-level &  &  & &  & \\
    \text{ }Mean-only & 0.019 & 0.059 & 0.866 & 0.151 & 0.678\\
    \text{ }Joint & 0.020 & 0.052 & 0.870 & 0.148 & 0.645\\
    Cluster-level &  &  &  &  & \\    
    \text{ }Beta-binomial & 0.013 & 0.053 & 0.997 & 0.064 & 0.745\\    
    \hline
    \end{tabular}
    }
    \caption{Results averaged over 500 experiments for Folate. Mean interval length(MIL), Coverage and  mean interval score(MIS) are all evaluated at the 90\% level.}
    \label{tab: Rwanda folate result}
\end{table}

\subsection{Estimation of ADM2 urban–rural proportions for Senegal}\label{apx:sen_urban}
ADM2-level urban–rural proportions are not reported for Senegal; only ADM1-level urban shares from the 2013 census are available. To construct ADM2 estimates, we combined WorldPop gridded population data with these ADM1 urban fractions. For each ADM1, we extracted census-year per-pixel population counts and determined a population threshold such that pixels with counts above this threshold accounted for the census-defined proportion of the ADM1 population classified as urban. Pixels were then labelled as urban or rural accordingly. Using both census-year and survey-year gridded population surfaces, we aggregated the urban and total population within each ADM2 to compute the ADM2-level urban share which were used in the cluster-level model applied to Senegal.

%%=============================================%%
%% For submissions to Nature Portfolio Journals %%
%% please use the heading ``Extended Data''.   %%
%%=============================================%%

%%=============================================================%%
%% Sample for another appendix section			       %%
%%=============================================================%%

%% \section{Example of another appendix section}\label{secA2}%
%% Appendices may be used for helpful, supporting or essential material that would otherwise 
%% clutter, break up or be distracting to the text. Appendices can consist of sections, figures, 
%% tables and equations etc.

\end{appendices}

%%===========================================================================================%%
%% If you are submitting to one of the Nature Portfolio journals, using the eJP submission   %%
%% system, please include the references within the manuscript file itself. You may do this  %%
%% by copying the reference list from your .bbl file, paste it into the main manuscript .tex %%
%% file, and delete the associated \verb+\bibliography+ commands.                            %%
%%===========================================================================================%%
\clearpage
\bibliography{sn-bibliography}% common bib file
%% if required, the content of .bbl file can be included here once bbl is generated
%%\input sn-article.bbl

\end{document}